\documentclass[conference]{IEEEtran}
\usepackage{amssymb}
\usepackage[cmex10]{amsmath} 
\usepackage{cite} 
\usepackage{url} 
\usepackage{multirow}
\usepackage{flushend}

\usepackage{ifpdf}
\ifCLASSINFOpdf
  \usepackage[pdftex]{graphicx}
  \DeclareGraphicsExtensions{.pdf}
\else
  \usepackage[dvips]{graphicx}
   \DeclareGraphicsExtensions{.eps}
\fi

\DeclareMathOperator*{\argmax}{arg\,max}
\begin{document}
\title{Signalling Storms in 3G Mobile Networks}
\author{\IEEEauthorblockN{Omer H. Abdelrahman and Erol Gelenbe}\\
\IEEEauthorblockA{
Department of Electrical \& Electronic Engineering\\
Imperial College, London SW7 2BT, UK\\
Email: \{o.abd06, e.gelenbe\}@imperial.ac.uk}}

\maketitle

\begin{abstract}
We review the characteristics of signalling storms that have been caused by certain common apps and recently observed in cellular networks, leading to system outages.  We then develop a mathematical model of a mobile user's signalling behaviour which focuses on the potential of causing such storms, and represent it by a large Markov chain. The analysis of this model allows us to determine the key parameters of mobile user device behaviour that can lead to signalling storms. We then identify  the parameter values that will lead to worst case load for the network itself in the presence of such storms. This leads to explicit results
regarding the manner in which individual mobile behaviour can cause overload conditions on the network and its signalling servers, and provides insight into how this may be avoided.
\end{abstract}

\section{Introduction}

Mobile networks are vulnerable to \emph{signalling attacks} which overload the control plane through traffic patterns that target the signalling procedures involved \cite{Serror2006,Enck2005,Lee2009,Ricciato2010}, by compromising a large number of mobile devices as in network Denial of Service (DoS) attacks \cite{Gelenbe2007,CACM-SAN} or from outside the mobile networks (e.g. the Internet). Similarly software and apps on mobile devices \cite{Viruses,Attack} can cause such disturbances through frequent traffic bursts. Such attackers can actively probe the network to infer the network's radio resource allocation policies \cite{Barbuzzi2008,Qian2010} and identify IP addresses in specific locations \cite{Qian2012}. Indeed, a review  of 180 cellular carriers around the world revealed that 51\% of them allow mobile devices to be probed from the Internet by either assigning public IP addresses to mobile devices or allowing IP spoofing or device-to-device probing within the network \cite{Wang2011,Qian2012}. Signalling attacks may also be launched in conjunction with the presence of crowds in well identified locations such as sports arenas or concert venues \cite{CAMWA}.

Signalling attacks are similar to  \emph{signalling storms} caused by poorly designed or misbehaving mobile apps that repeatedly establish and tear down data connections \cite{NSN2011}, generating large amounts of signalling that may crash the network. Such signalling storms are a serious threat to the availability and security of cellular networks.  While flash crowds last for a short time during  special occasions such as New Year's Eve, signalling storms are unpredictable and tend to persist until the underlying problem is identified and corrected. This has prompted the industry to promote best practices for developing ``network-friendly'' mobile apps \cite{GSMA2012,Jiantao2012}.

\subsection{Signalling Storms}

Perhaps one of the most important features of smart phones and tablets is the ``always-on'' connectivity, which enables users to receive push messages, e.g. to notify of an incoming message or VoIP call. This is maintained by having the mobile device send periodic keep-alive messages to a cloud server. However, if for any reason the cloud service becomes unavailable, then the mobile device will attempt to reconnect more frequently generating signalling loads up to 20 times more than normal as reported in recent incidents \cite{Nokia2013}. In 2012 a Japanese mobile operator suffered a major outage \cite{Storm2012} due to a VoIP app that constantly polls the network even when users are inactive. In another incident \cite{Corner2011} the launch of a free version of a popular game on Android caused signalling overload in a large network due to frequent advertisements shown within the app. Also, many mobile carriers have reported \cite{Arbor2012} outages or performance issues caused by non-malicious but misbehaving apps, yet the majority of those affected followed a reactive approach to identify and mitigate the problem.

Signalling storms could also occur as a byproduct of large scale malware infections \cite{Ricciato06}, such as botnets, which target mobile users rather than networks. A recent report by Kaspersky \cite{Kaspersky2013} revealed that the most frequently detected malware threats affecting Android OS are (i) SMS trojans which send costly messages without users' consent, (ii) adware which displays unwanted advertisements, and (iii) root exploits which allow the installation of other malware or the device to become part of a botnet. A sufficiently large number of users within a single network falling victims to such attacks, which involve frequent communications, could have a devastating impact on the control plane of the network.

The purpose of this paper is to analyse the effect of signalling storms, as well as of signalling attacks, and analyse in particular the manner in which such attacks can cause maximum damage to the radio and core networks. The approach we take is based on the development of a mathematical model of user signalling behaviour from which we derive some useful analytical results. While the literature \cite{Haverinen2007,Yeh2009,Schwartz2013} has focused on analysing signalling behaviour from an energy consumption perspective, we hope that this work can offer to mobile operators a greater understanding of bottlenecks and vulnerabilities in the radio signalling system, so that network parameters may be modified so as to mitigate for those effects that lead to network outages \cite{nemesys1,nemesys2}.

\section{Modelling Signalling of a Single User}

In the context of UMTS networks, bandwidth is managed by the {\em radio resource control} (RRC) protocol which associates a state machine with each {\em user equipment} (UE). There are typically four RRC states, in order of increasing energy consumption: IDLE, Paging Channel (cell\_PCH), low bandwidth Forward Access Channel (cell\_FACH), and high bandwidth Dedicated Channel (cell\_DCH). We will refer hereafter to state cell\_X as X. State promotions are triggered by uplink (UL) and downlink (DL) transmissions, and the move to FACH or DCH is determined by the size of the {\em radio link control} (RLC) buffer of the UE: if at any time the buffer exceeds a certain threshold in either direction, the state will be promoted to DCH. State demotions are triggered by inactivity timers.

Consider a UE that transitions from IDLE or dormant $D$ to FACH, perhaps later to DCH, and then sometimes directly from $D$ to DCH. We will let $\lambda_L$ and $\lambda_H$ be the rates at which low and high bandwidth calls\footnote{A call refers to any UL/DL activity, e.g. data session, location update, etc.} are normally made, and $L$ and $H$ be the corresponding states when the call is actually taking place in the sense that it is using the bandwidth of FACH and DCH. Furthermore, we will denote by $\eta$ the state when a low bandwidth request is handled while the mobile is in DCH.

At the end of normal usage the call will transition from $L$ to $\ell$ or from $H,\eta$ to $h$, where $\ell$ and $h$ are the states when the UE is not using the bandwidth of FACH and DCH respectively; thus, $\{L,\ell\}\in$ FACH and $\{H,\eta,h\}\in$ DCH. We denote the rates at which low and high bandwidth calls terminate by $\mu_L$ and $\mu_H$. Since the amount of traffic exchanged in states $L$ and $\eta$ is usually very small (otherwise it will trigger a transition to $H$), we assume that their durations are independent but stochastically identical.

If the UE does not start a new session for some time, it will be demoted from $h$ to $\ell$ or from $\ell$ to PCH which we denote by $P$. The UE will then return from $P$ to $D$ after another inactivity timer; however, because the mobile is not allowed to communicate in the $P$ state, it will first move to FACH, release all signalling connections, and finally move to $D$. Let $\tau_H$, $\tau_L$ and $\tau_P$ be the time-out rates in states $h, \ell$ and $P$, respectively.

We are considering signalling attacks (or misbehaving apps) which falsely induce the mobile to go from $D, P$ to FACH or DCH, or from FACH to DCH, without the user actually having any usage for this request.
The rates related to these malicious transitions will be denoted $\alpha_L$ and $\alpha_H$. Since in these cases a transition to an actual bandwidth usage state does not take place, unless the user starts a new session, the timers will demote the state of the UE. Consequently, the attack results in the usage of network resources both by the computation and state transitions that occur for call handling, and through bandwidth reservation that remains unutilised.

In summary, the state of the UE at time $t$ is described by the variable $s(t)\in \{\mathcal{N},\mathcal{A},\mathcal{S(N)},\mathcal{A(N)}\}$ where:
\begin{itemize}
\item $\mathcal{N}=\{D,P,\ell,L,h,\eta,H\}$ represent the states occupied by the UE during or after a ``normal'' call.
\item $\mathcal{A}=\{\ell_A,h_A,\eta_A\}$ are similar to $\{\ell,h,\eta\}$ but forced by malicious traffic. Note that a transition to state $\eta_A$ happens either from $L$ because of an attack that forces an ongoing low bandwidth call to communicate over DCH, or from $h_A$ because of a new normal low bandwidth call that could have been handled through FACH.
\item $\mathcal{S(N)}$ and $\mathcal{S(A)}$ are the signalling states for respectively normal and attack conditions, which capture the non-negligible overhead needed in order to establish and release network resources during state promotions and demotions. We denote by $\sigma_{XY}^{-1}$ the average transition delay from state $X$ to $Y$, where $X,Y\in \{D,P,L,H\}$ and the subscripts $L$ and $H$ are used here to represent both normal and attack states in FACH and DCH.
\end{itemize}
\begin{figure}[]\centering%
   \includegraphics[width=0.49\textwidth]{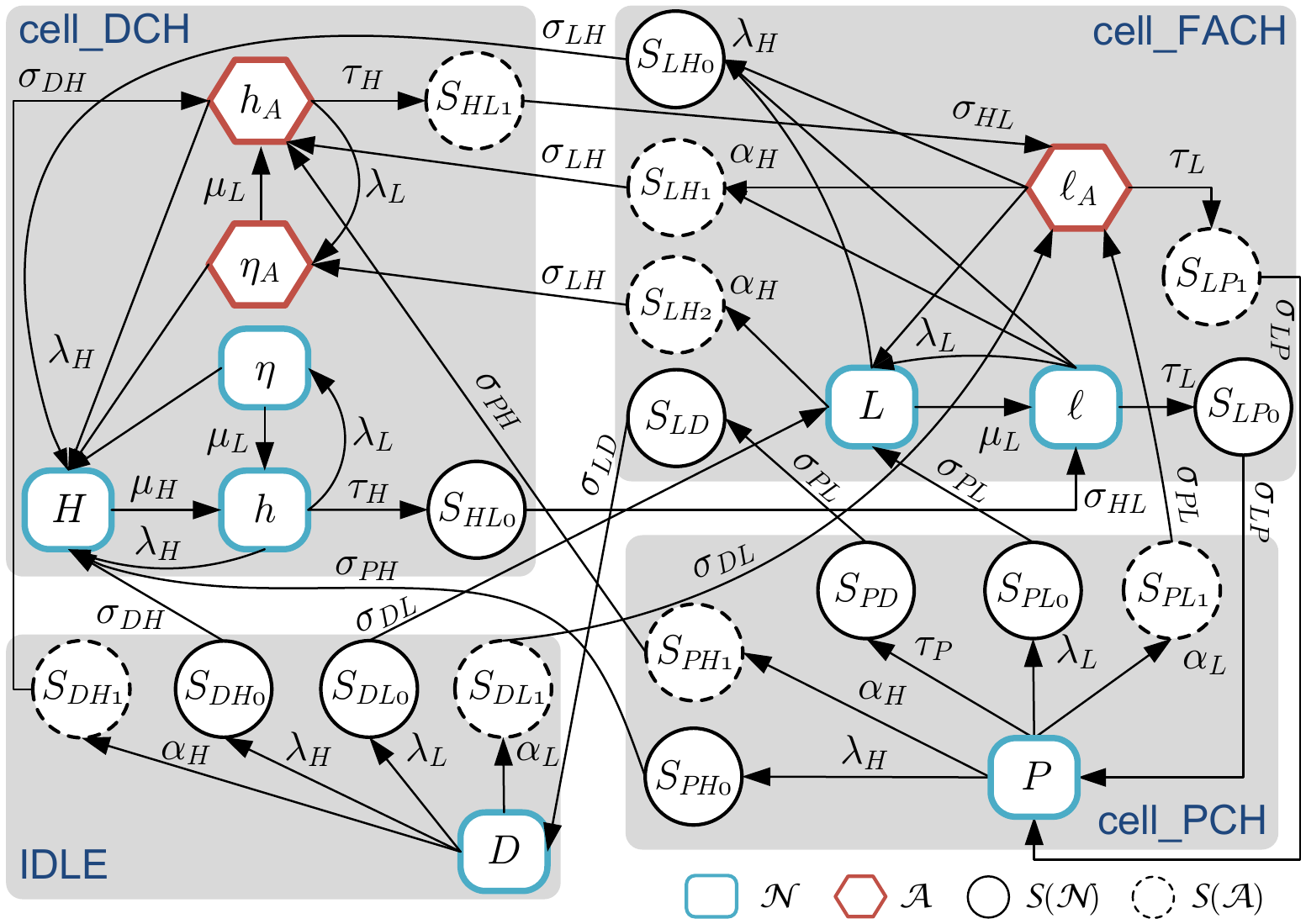}
   \caption{The Markov model of a single user.} \label{fig-model}
\end{figure}
Fig.~\ref{fig-model} shows the different states, signalling phases and transitions of the Markov model. The stationary equations for the states in $\mathcal{N}$ are given by:
\begin{align*}
&\pi(D)[\lambda_H + \alpha_H + \lambda_L + \alpha_L]  =  \pi(P) \tau_P,\\
&\pi(P)[\lambda_H + \alpha_H + \lambda_L + \alpha_L + \tau_P]  =  [\pi(\ell) + \pi(\ell_A)] \tau_L,\\
&\pi(\ell) [\lambda_H + \alpha_H + \lambda_L + \tau_L] =  \pi(L) \mu_L + \pi(h) \tau_H, \\
&\pi(L) [\lambda_H + \alpha_H  + \mu_L] = [\pi(D) + \pi(P) + \pi(\ell) + \pi(\ell_A)] \lambda_L, \\
&\pi(h)[\lambda_H + \lambda_L + \tau_H]  = \pi(H) \mu_H  + \pi(\eta) \mu_L,  \\
&\pi(\eta)[\lambda_H + \mu_L]  =  \pi(h)\lambda_L, \\
&\pi(H)\mu_H  =  \sum_{s\in \{ \mathcal{N},\mathcal{A} \}, s \neq H}\pi(s) \lambda_H,
\end{align*}
while the equations for the attack states $\mathcal{A}$ are:
\begin{align*}\nonumber
&\pi(\ell_A) [\lambda_H + \alpha_H + \lambda_L + \tau_L] = [\pi(D)+\pi(P)]  \alpha_L +  \pi(h_A) \tau_H,\\\nonumber
&\pi(h_A)[\lambda_H + \tau_H + \lambda_L ]  =  \sum_{s\in \{D,P,\ell,\ell_A\}}\pi(s) \alpha_H + \pi(\eta_A) \mu_L,  \\
&\pi(\eta_A)[\lambda_H + \mu_L]  =  \pi(h_A) \lambda_L + \pi(L) \alpha_H.
\end{align*}
We can express the normalisation condition as a weighted sum of the probabilities of the states $\{\mathcal{N},\mathcal{A}\}$, i.e.  $1=\sum_{s \in \{\mathcal{N},\mathcal{A}\}} \pi(s)w_s$ or:
\begin{align}\nonumber
&1 =    \underbrace{\pi(D)[1 + \frac{\Lambda_H}{\sigma_{DH}} + \frac{\Lambda_L}{\sigma_{DL}}]}_{\Pr[\text{user in IDLE}]}\\\nonumber
& + \underbrace{\pi(P)[1 + \frac{\Lambda_H}{\sigma_{PH}} + \frac{\Lambda_L}{\sigma_{PL}} +  \tau_P (\frac{1}{\sigma_{PL}}}_{\Pr[\text{user in PCH}]} + \frac{1}{\sigma_{LD}})]\\\nonumber
& + (\pi(\ell)+\pi(\ell_A))[1+ \frac{\Lambda_H}{\sigma_{LH}} + \frac{\tau_L}{\sigma_{LP}}] + \pi(L)[1+ \frac{\Lambda_H}{\sigma_{LH}}]\\\label{norm}
& + \underbrace{(\pi(h)+\pi(h_A))[1+\frac{\tau_H}{\sigma_{HL}}]  + \pi(\eta) + \pi(\eta_A)+\pi(H)}_{\Pr[\text{user in DCH}]}
\end{align}
with $\Lambda_H = \lambda_H + \alpha_H$ and $\Lambda_L = \lambda_L + \alpha_L$. Writing $\Lambda  = \Lambda_L + \Lambda_H$, $q_L =\frac{\lambda_L}{\lambda_H + \mu_L}$, $\rho_L = \frac{\lambda_L}{\Lambda_H + \mu_L}$, and $q_H = \frac{\lambda_H}{\mu_H}$, the solution to the above set of equations becomes:
\begin{align*}
\pi(D)=~& \frac{\tau_P \tau_L}{(\Lambda + \tau_P)(\Lambda + \tau_L)} ~ G,\\
\pi(P)=~& \frac{\Lambda \tau_L }{(\Lambda + \tau_P)(\Lambda + \tau_L)}~ G,\\
\pi(L)=~& \rho_L ~G,\\
\pi(H)=~& q_H [\frac{q_L\rho_L\alpha_H }{\lambda_L} + (1+\rho_L)(\frac{\Lambda_H}{\tau_H}[1+ q_L] + 1 ) ] ~G,\\
\pi(h)=~& \frac{\mu_H}{\lambda_H [1+q_L] + \tau_H } ~ \pi(H), \\
\pi(\eta)=~& q_L ~ \pi(h),\\
\pi(\ell)=~& \frac{1}{\Lambda_H + \lambda_L + \tau_L} [\mu_L \rho_L G + \frac{\mu_H \tau_H\pi(H)}{\lambda_H [1+q_L] + \tau_H} ],\\
\pi(h_A)=~& \frac{\alpha_H}{\lambda_H [1+q_L] + \tau_H}[1+ \frac{q_L \rho_L\mu_L}{\lambda_L}]~  G,\\
\pi(\eta_A)=~& \frac{\alpha_Hq_L}{\lambda_H [1+q_L] + \tau_H} [ 1 + \frac{\lambda_H + \tau_H + \lambda_L}{\Lambda_H + \mu_L}]G,\\
\pi(\ell_A)=~& \frac{1}{\Lambda_H + \lambda_L + \tau_L}[\frac{\alpha_L \tau_L}{\Lambda + \tau_L} + \frac{\alpha_H \tau_H (1+\frac{q_L \rho_L\mu_L}{\lambda_L})}{\lambda_H [1+q_L] + \tau_H}]G,
\end{align*}
where $G$ can be obtained from \eqref{norm} yielding:
\begin{align*}
&G^{-1}=  [1+\rho_L][q_H  + \frac{\Lambda_H}{\tau_H} \{(1+ q_L) (1 + q_H)+ w_h - 1\} ]+\\
& \frac{\frac{\tau_L}{\Lambda + \tau_P} [\tau_Pw_D +\Lambda w_P] + \Lambda w_\ell}{\Lambda + \tau_L}  + \rho_L [ w_L + \frac{q_L}{\lambda_L}(1+ q_H)\alpha_H].
\end{align*}

\subsection{Signalling Load on the RNC and SGSN}

Let $n_{XY}$ denote the number of signalling messages sent or received by the {\em radio network controller} (RNC) when a transition occurs from state $X$ to state $Y$, then the signalling rate generated by a single user due to both normal and malicious traffic can be computed as:
\begin{align}\nonumber
\gamma_{r} ~=~ & \pi(D)[\Lambda_H n_{DH} + \Lambda_L n_{DL}] + \pi(P)[\Lambda_H n_{PH}  + \Lambda_L n_{PL}] \\\nonumber
               & + [\pi(\ell) + \pi(\ell_A) + \pi(L)] \Lambda_H n_{LH}  \\\nonumber
               & + [\pi(h) + \pi(h_A)] \tau_H n_{HL}  \\\nonumber
               & + [\pi(\ell) + \pi(\ell_A)] \tau_L \{n_{LP} \mathbf{1_{L\to P}} + n_{LD} \mathbf{1_{L\to D}}\} \\\label{radio-rate}
               & + \pi(P)\tau_P n_{PD}\mathbf{1_{L\to P}},
\end{align}
where the characteristic function $\mathbf{1_{X\to Y}}$ takes the value 1 if the transition $X\to Y$ is implemented and 0 otherwise. Note that the mobile network operator may not use PCH state,  e.g. when the vendor does not support it or it is disabled in order to extend the battery life of mobile devices. In this case, $\sigma_{PL},~\sigma_{LP}$ and $\tau_P$ are set to $\infty$ so that the user is moved directly from FACH to IDLE after an inactivity timer.

On the other hand, the core network is more protected from signalling attacks since only transitions to/from state $D$ trigger signalling with the core. Let $m_{XY}\leq n_{XY}$ be the number of control plane messages exchanged between the RNC and the {\em serving GPRS support node} (SGSN) during such transitions, then the signalling load on the core network from a single user becomes:
\begin{align}\nonumber
\gamma_{c} ~=~& \pi(D)[\Lambda_H m_{DH} + \Lambda_L m_{DL}] + \pi(P) \tau_P m_{PD}\mathbf{1_{L\to P}}\\\label{core-rate}
              & +  [\pi(\ell) + \pi(\ell_A)] \tau_L m_{LD} \mathbf{1_{L\to D}}.
\end{align}

Table~\ref{table-param} summarises the state transition model along with parameter values used in the numerical results: (i) the number of signalling messages exchanged during state transitions are obtained from the UMTS standards documentation, and can also be found in the literature (e.g. \cite{GSMA2011}); (ii) typical values for the inactivity timers $\tau_H^{-1}$ and $\tau_L^{-1}$ are in the range $2-10$~seconds, while $\tau_P^{-1}$ should be significantly longer (in the order of minutes); and (iii) the average transition times are assumed to be proportional to the number of signalling messages involved, and normalised with respect to the transition IDLE $\to$ DCH which is assumed to take 1~second.

\begin{table}[t!]\caption{Network parameters}\label{table-param}
\begin{tabular}{p{2cm}|p{3.85cm}|p{0.3cm}|p{0.35cm}|p{0.3cm}}\hline
  \textbf{Transition} & \textbf{Triggering Event} & $\hspace{-0.17cm}\mathbf{n_{XY}}$ & $\hspace{-0.17cm}\mathbf{m_{XY}}$ & $\hspace{-0.1cm}\mathbf{\sigma_{XY}^{-1}}$ \\\hline\hline
  IDLE $\to$ FACH  & \multirow{3}{1.5in}{Low bandwidth UL/DL traffic (e.g. location update, keep-alive messages)} & 15 & 5  & 0.75\\
  PCH~ $\to$ FACH  &                                        & ~3  & -- & 0.15   \\
  &&&&\\\hline
  IDLE $\to$ DCH   & \multirow{3}{1.5in}{High bandwidth UL/DL traffic (e.g. VoIP calls, video streaming, web browsing)}& 20 & 5  & 1.0   \\
  PCH~ $\to$ DCH   &                                        & 10  & -- & 0.5   \\
  FACH $\to$ DCH   &                                        & ~7  & -- & 0.35   \\\hline
  DCH $\to$ FACH   & inactivity timer $\tau_H^{-1} = 2-10$s   & ~5  & -- & 0.25  \\
  FACH $\to$ PCH   & inactivity timer $\tau_L^{-1} = 2-10$s  & ~2  & -- & 0.1 \\
  PCH ${\tiny \xrightarrow{\text{FACH}}}$ IDLE    & inactivity timer $\tau_p^{-1} = 5-20$min & ~6  & 2  & 0.3  \\\hline
\end{tabular}
\end{table}

\section{Maximising the Impact of an Attack}

\begin{figure}[t!]\centering%
   \includegraphics[width=0.49\textwidth]{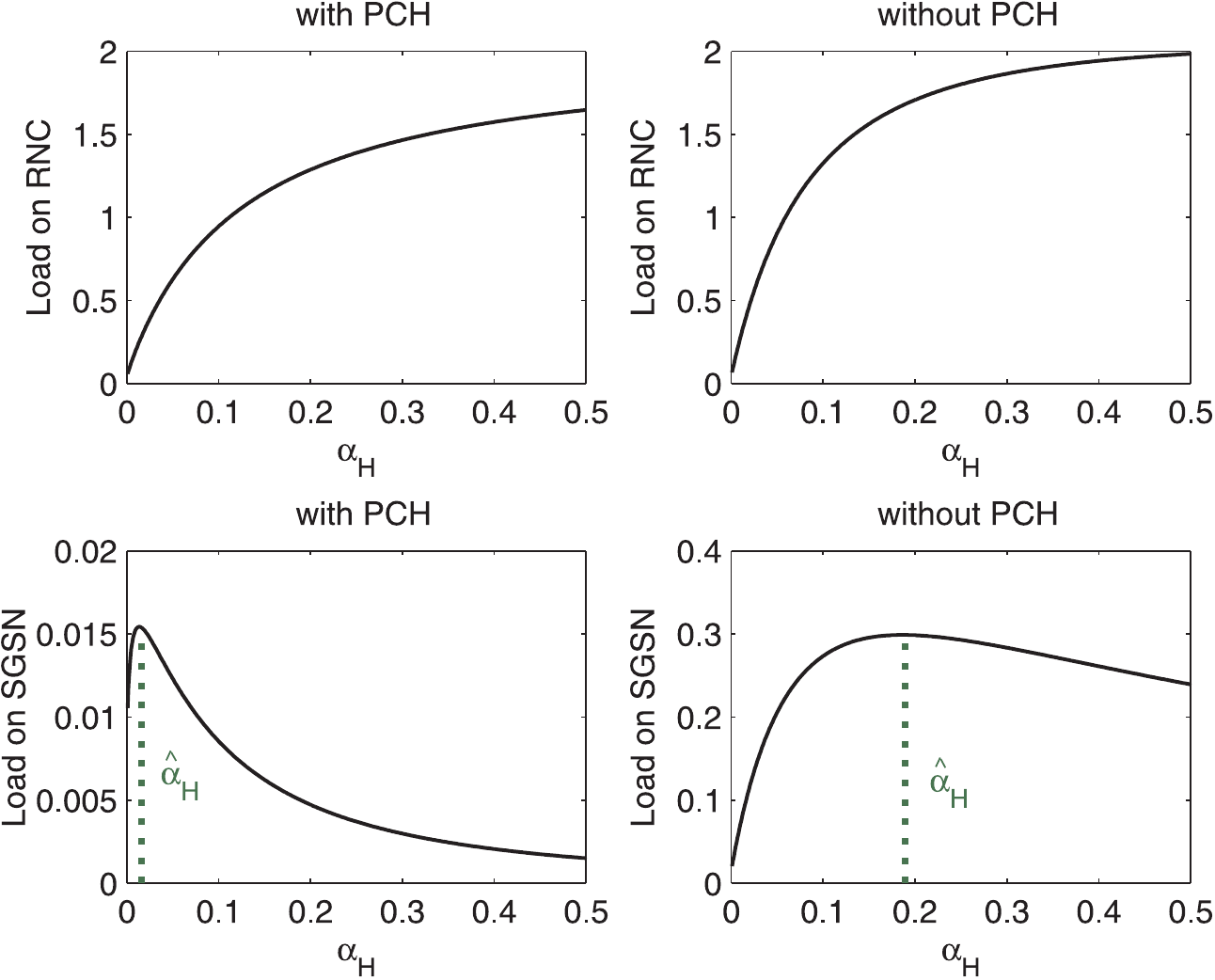}
   \caption{The average signalling load ($msg/s$) on RNC and SGSN with and without PCH state versus attack rate $\alpha_H$ when $\alpha_L = 0$, normal traffic is characterised by $\lambda_L^{-1} = 600, \mu_L^{-1} = 5, \lambda_H^{-1} = 1800, \mu_H^{-1} = 120$, and using the parameters of Table~\ref{table-param} with $\tau_H^{-1} = \tau_L^{-1} = 5$s and $\tau_P^{-1} = 5$min. $\hat{\alpha}_H$ in \eqref{approx} provides a good estimate of the optimum attack rate.} \label{fig-gamma}
\end{figure}

If an attacker succeeds in inferring the radio network configuration parameters (e.g. through active probing \cite{Barbuzzi2008,Qian2010,Qian2012}), then it is easy to monitor the user's behaviour in order to estimate $\lambda_L, \lambda_H, \mu_L$ and $\mu_H$. The attacker can then maximise the impact on the radio or core network by choosing the rate of malicious traffic bursts $\alpha_L$ and $\alpha_H$ so as to maximise \eqref{radio-rate} or \eqref{core-rate}. This is illustrated in Fig.~\ref{fig-gamma} where we plot the average rate of signalling messages that a misbehaving user generates on the RNC and SGSN assuming $\alpha_L = 0$ and different values of $\alpha_H$. The results indicate that there is indeed an optimum value of $\alpha_H$ which maximises the load on the {\em core network}, while the load on the radio network increases monotonically with the attack rate up to a maximum level.

The effect of PCH state is also examined in Fig.~\ref{fig-gamma} showing a significant reduction (about 95\%) in the amount of control plane traffic reaching the core network as compared to the case where the user is moved directly from FACH to IDLE. In fact, as the value of the timer $\tau_P^{-1}$ gets larger, an attacker would find it extremely difficult to overwhelm the SGSN with signalling load unless a very large number of UEs are compromised. This feature also results in up to 30\% drop in the amount of signalling load traversing the radio network.

\subsection{Radio Network}\label{sec:storm_radio}

Numerical investigations suggest that the load on the radio network increases with the frequency of the malicious bursts up to a maximum level reached when {\em either} $\alpha_H$ or $\alpha_L$ tends to infinity, depending on the parameters of the network as well as the user's traffic characteristics. If PCH is enabled then the attacker could either induce the transition FACH $\to$ DCH as soon as the channel is released, or take a two-step approach to first move from PCH to FACH immediately after the timer $\tau_L^{-1}$ expires then trigger another transition to DCH some time later. Note that any other attack policy would be slowed down by the long timer $\tau_P^{-1}$ and thus would not succeed in creating a more severe impact. To investigate both policies, let us set $\alpha_L \to \infty$ so that the transition PCH $\to$ FACH is triggered repeatedly, creating a load on the radio network given by:
\begin{align*}
\gamma_r ~=~&  \frac{n_{LH} + n_{HL}}{\theta_{LH} + \frac{\tau_L(\Lambda_H + \mu_L)}{\Lambda_H (\Lambda_H + \lambda_L + \mu_L)}~\theta_{PL}} \\
& +\frac{n_{PL} + n_{LP}}{\theta_{PL} + \frac{\Lambda_H (\Lambda_H + \lambda_L + \mu_L)}{\tau_L(\Lambda_H + \mu_L)}~\theta_{LH}},\qquad \alpha_L \to \infty,
\end{align*}
where $\theta_{XY} = \sigma_{XY}^{-1}  + (1 + q_L)(1 + q_H)\tau_Y^{-1}+\sigma_{YX}^{-1}$. Now if we maximise the above expression with respect to $\alpha_H$, we obtain the following interesting result:
\begin{equation}
(\alpha_L^*,\alpha_H^*) = \left\{
\begin{array}{ll}
  (\infty, 0),        & \text{if}~~\frac{n_{LH} + n_{HL}}{\theta_{LH}}\leq \frac{n_{PL} + n_{LP}}{\theta_{PL}},\\
  (0,\infty),   & \text{otherwise.}
\end{array}\right.
\end{equation}
Therefore, the load on the radio network can be maximised through low (resp. high) bandwidth bursts that repeatedly induce the transition PCH $\to$ FACH (resp. FACH $\to$ DCH) if the condition $[n_{LH} + n_{HL}]\theta_{LH}^{-1}\leq [n_{PL} + n_{LP}]\theta_{PL}^{-1}$ is (resp. is not) satisfied. When PCH state is not used, we obtain similar results, but the attack is maximised by continuously triggering IDLE $\to$ FACH or FACH $\to$ DCH depending on whether the condition $[n_{LH} + n_{HL}]\theta_{LH}^{-1}\leq [n_{DL} + n_{LD}]\theta_{DL}^{-1}$ is satisfied or not, respectively. The worst case load on the RNC is then:
\begin{equation*}
\gamma_r^* = \max\left[\frac{n_{XL} + n_{LX}}{\theta_{XL}},\frac{n_{LH} + n_{HL}}{\theta_{LH}} \right],
\end{equation*}
$X=D$ or $P$ depending on which transition $L\to X$ is used.

\subsection{Core Network}\label{sec:storm_core}

\begin{figure}[t!]\centering%
   \includegraphics[width=0.38\textwidth]{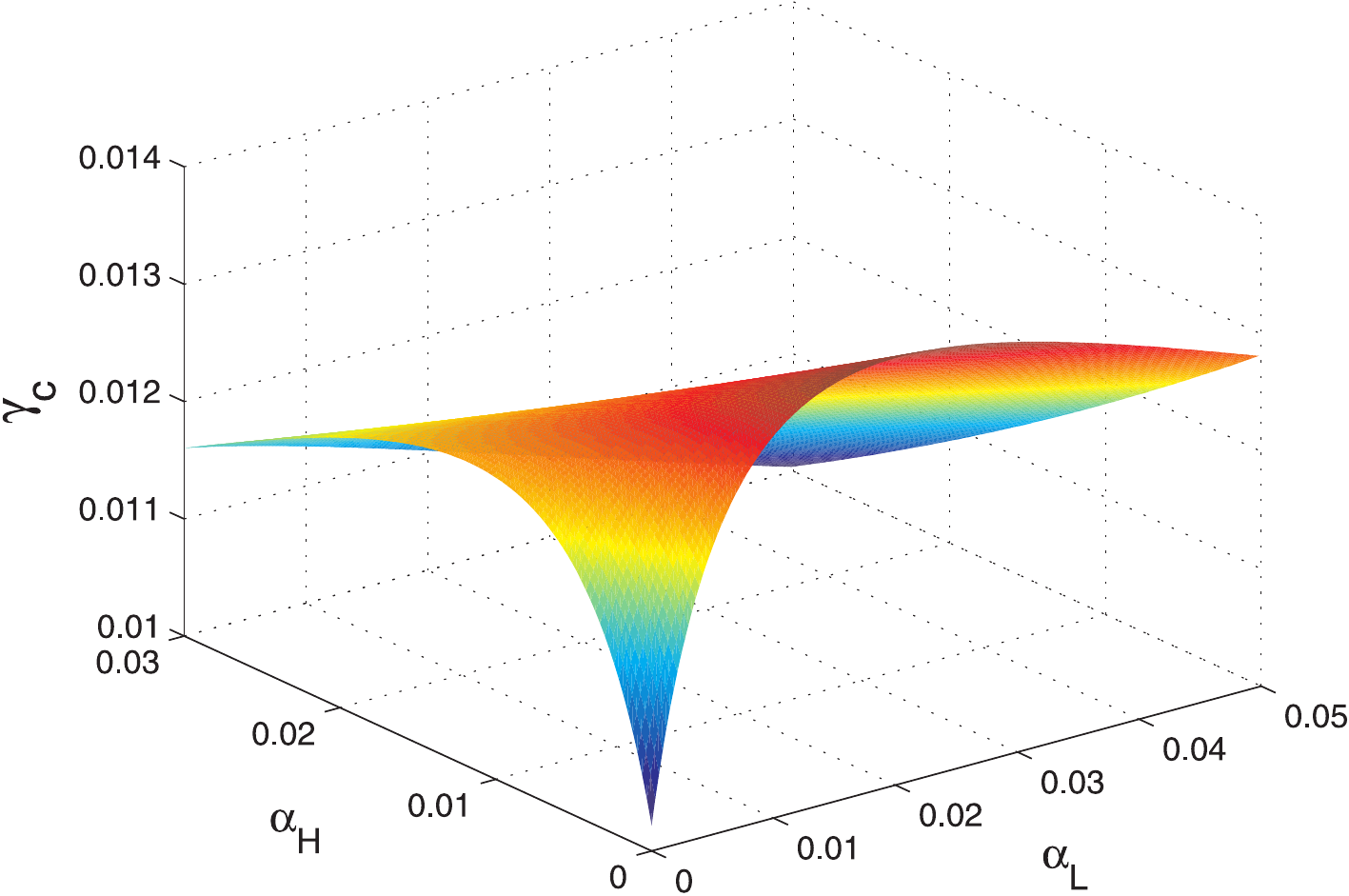}
    \caption{The average signalling load ($msg/s$) on SGSN versus the attack rates $\alpha_H, \alpha_L$, when normal traffic profile is $\lambda_L^{-1} = 300, \mu_L^{-1} = 5, \lambda_H^{-1} = 600, \mu_H^{-1} = 180$, and the timers are $\tau_H^{-1} = \tau_L^{-1} =5$s and $\tau_P^{-1} = 5$~min.} \label{fig-gc}
\end{figure}

Signalling between the UE and core network happens for a number of different reasons, but with respect to the RRC state machine, it usually occurs when the UE moves from/to the IDLE state. The attack against the core network can then be launched more effectively by causing a transition to FACH, rather than DCH, immediately after the user becomes IDLE so as to avoid the timer $\tau_H^{-1}$ and the associated demotion delay. Thus, optimally $\alpha_H^* =0$, and the attack rate that maximises the load on the core network can be shown to be:
\begin{equation}\label{alpha-Lc}
\alpha_L^* =  \sqrt{c^2 +  \frac{b-ca}{\theta_{PLH}} } -c -\lambda_L ,
\end{equation}
where:
\begin{align*}
&\theta_{PLH} = \theta_{PL} + (1+q_L)\tau_L^{-1}\lambda_H \theta_{LH},\\
&a =   \lambda_H [2\theta_{PLH} + \sigma_{DH}^{-1} - \sigma_{PL}^{-1} - \sigma_{LH}^{-1}] \\
    &~~ + \tau_P [\theta_{PLH} + \sigma_{DL}^{-1} + \sigma_{LD}^{-1}]+(1 + q_L)(1 + q_H + \lambda_H \theta_{LH}),\\
&b = \lambda_H^2 [\theta_{PLH} + \sigma_{PH}^{-1} - \sigma_{PL}^{-1} - \sigma_{LH}^{-1}]\\
    &~~ + \lambda_H \tau_P [\theta_{PLH} + \sigma_{DH}^{-1} + \sigma_{LD}^{-1} - \sigma_{LH}^{-1}]\\
    &~~ +(\lambda_H + \tau_P)(1 + q_L)(1 + q_H + \lambda_H \theta_{LH}),\\
&c =  \lambda_H ~ \frac{m_{DH} + m_{PD}}{m_{DL} + m_{PD}}.
\end{align*}
Obviously, the attack is worst when there is no background high bandwidth user traffic, in which case we end up with:
\begin{equation*}
\alpha_L^* = \sqrt{\frac{\tau_P [1+\frac{\lambda_L}{\mu_L}]}{\theta_{PL}}} - \lambda_L, \qquad \lambda_H = 0
\end{equation*}
and consequently the maximum possible load that an attacker can impose on the SGSN is:
\begin{align*}
\gamma_c^* = \frac{m_{DL} + m_{PD}}{\sigma_{DL}^{-1}+ \sigma_{LP}^{-1}+ \sigma_{PL}^{-1}+ \sigma_{LD}^{-1}
+ (1+\frac{\lambda_L}{\mu_L})(\frac{1}{\tau_L} +\frac{1}{\tau_P} +  \frac{2}{\Lambda_L^*})}
\end{align*}
with $\Lambda_L^* = \alpha_L^* + \lambda_L$. When $\tau_P \to \infty$, we get the intuitive result $\alpha_L^* = \infty$, i.e. the attacker should send a low-bandwidth traffic burst as soon as the timer $\tau_L^{-1}$ expires, leading to $\gamma_c^* =[m_{DL} + m_{LD}]\theta_{DLH}^{-1}$ where $\theta_{DLH} = \theta_{DL} + (1+q_L)\tau_L^{-1}\lambda_H \theta_{LH}$. In Fig.~\ref{fig-gc} we plot $\gamma_c$ versus the attack rates, and the numerical results indicate that $(\alpha_L^*,\alpha_H^*) = (0.02, 0)$ which coincide with the prediction of \eqref{alpha-Lc}.

In practice, however, mounting an attack based solely on low bandwidth bursts may not be feasible. To begin with, it may be difficult to accurately estimate the RLC buffer's thresholds which determine whether a session will be handled through the low or high speed channel, and also the thresholds could differ from one RNC to another. Furthermore, many operators choose to move users directly into DCH or use very small RLC thresholds such that even keep-alive messages are sent over the high speed channel \cite{Qian2010}. Thus, a more practical approach for an attacker is to assume that the majority of data transmissions are handled through DCH, and in turn compute an attack rate $\hat{\alpha}_H$ that maximises the load on the SGSN under such circumstances, i.e.:
\begin{equation*}
\hat{\alpha}_H   = \argmax_{\alpha_H} \quad \gamma_c, \qquad \text{when}\quad \Lambda_H >> \Lambda_L,
\end{equation*}
yielding:
\begin{align}\nonumber
\hat{\alpha}_H ~=~& \sqrt[3]{-\frac{B}{2} + \sqrt{\frac{B^2}{4} + \frac{A^3}{27}}}+ \sqrt[3]{-\frac{B}{2} - \sqrt{\frac{B^2}{4} + \frac{A^3}{27}}}\\\label{approx}
                  & ~~ - \frac{b}{6a} - \lambda_H,
\end{align}
where:
\begin{align*}
A &= -\frac{b^2}{12a^2}, \quad B = \frac{b^3}{108a^3} - \frac{c}{2a}, \\
a &= \sigma_{LH}^{-1}  +  [1 + \frac{\lambda_H}{\mu_H}] \tau_H^{-1}+ \sigma_{HL}^{-1}, \\
b &= \tau_L (\sigma_{PH}^{-1} + [1 + \frac{\lambda_H}{\mu_H}][\tau_H^{-1}+\tau_L^{-1}]+ \sigma_{HL}^{-1}+ \sigma_{LP}^{-1}) + \tau_P a , \\
c &= \tau_L \tau_P  [1 + \frac{\lambda_H}{\mu_H}].
\end{align*}
When PCH is disabled, we have:
\begin{equation*}
\hat{\alpha}_H=\sqrt{\frac{\tau_L [1+\frac{\lambda_H}{\mu_H}]}{\sigma_{LH}^{-1}  + [1 + \frac{\lambda_H}{\mu_H}]\tau_H^{-1} +  \sigma_{HL}^{-1} } }- \lambda_H
\end{equation*}
and the resulting load on the SGSN becomes:
\begin{equation*}
\hat{\gamma}_c = \frac{m_{DH} + m_{LD}}{\sigma_{DH}^{-1}+ \sigma_{HL}^{-1}+ \sigma_{LD}^{-1}+ (1+\frac{\lambda_H}{\mu_H})(\frac{1}{\tau_H}  +\frac{1}{\tau_L} +  \frac{2}{\hat{\alpha}_H+\lambda_H})}.
\end{equation*}
Fig.~\ref{fig-gamma} shows that $\hat{\alpha}_H$ provides a good estimate of the optimum value $\alpha_H^*$ even when $\lambda_L > \lambda_H$.

\begin{figure}[t!]\centering%
 \includegraphics[width=0.49\textwidth]{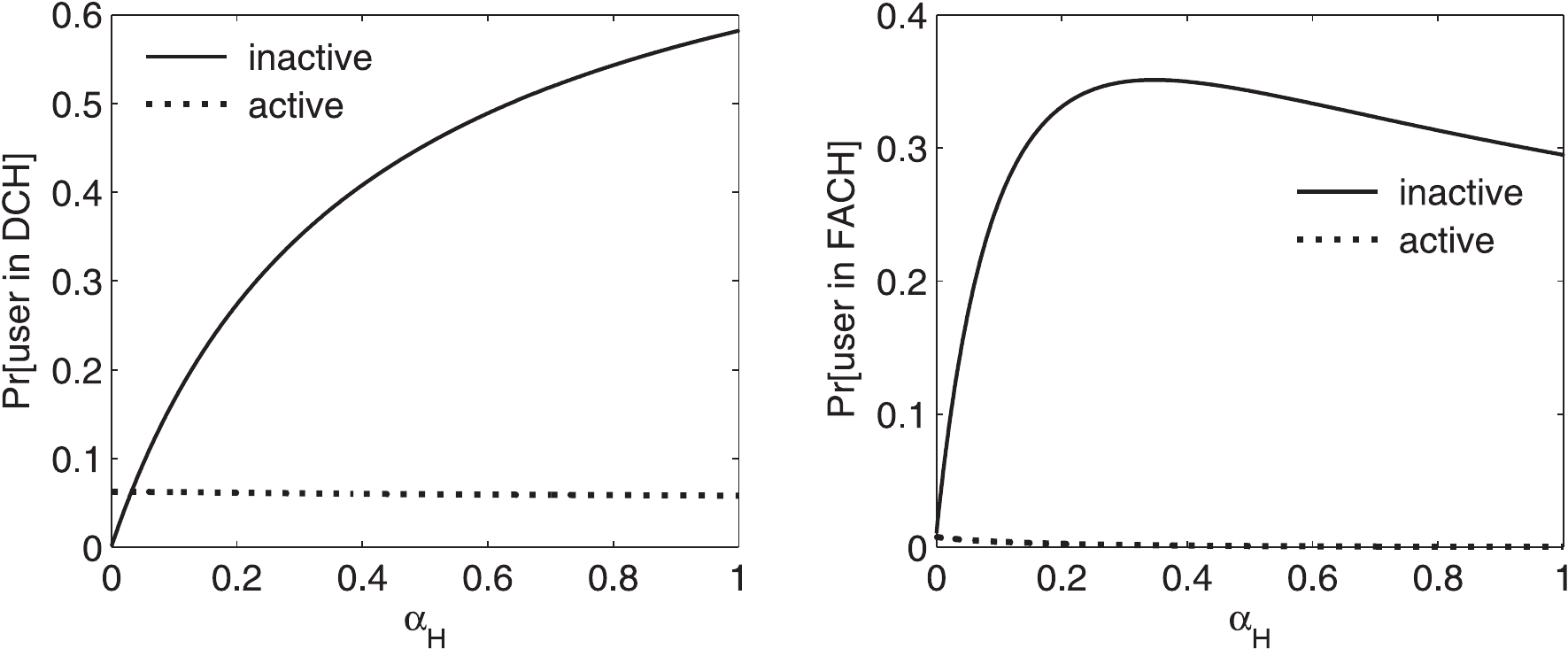}
   \caption{The fraction of time the UE spends in DCH and FACH waiting for a timer or state transition (solid line) and while using the bandwidth (dotted line) as a function of $\alpha_H$, when $\alpha_L = 0, \lambda_L^{-1} = 600, \mu_L^{-1} = 5, \lambda_H^{-1} = 1800, \mu_H^{-1} = 120, \tau_H^{-1} = 2$s, $\tau_L^{-1} = 5$s and $\tau_P^{-1} = 10$~min. Large inactive times indicate anomalous signalling behaviour.} \label{fig-util}
\end{figure}

Fig.~\ref{fig-util} illustrates the manner in which the frequency of malicious traffic bursts affects signalling overhead as well as the {\em tail} which is the time the UE spends in FACH or DCH waiting for a time-out to expire. During these inactive periods, the mobile wastes considerable radio resources in the network as well as its own limited battery energy. As the attack rate increases, the proportion of time the UE remains inactive in either FACH or DCH also increases, while its average data volume is almost constant. This observation could be used by anomaly detection techniques to distinguish between normal ``heavy'' users and attackers: the former can be recognised by their low inactive times, while the latter can be detected by frequent connection attempts and low data volume.

\begin{figure}[t!]\centering%
  \includegraphics[width=0.49\textwidth]{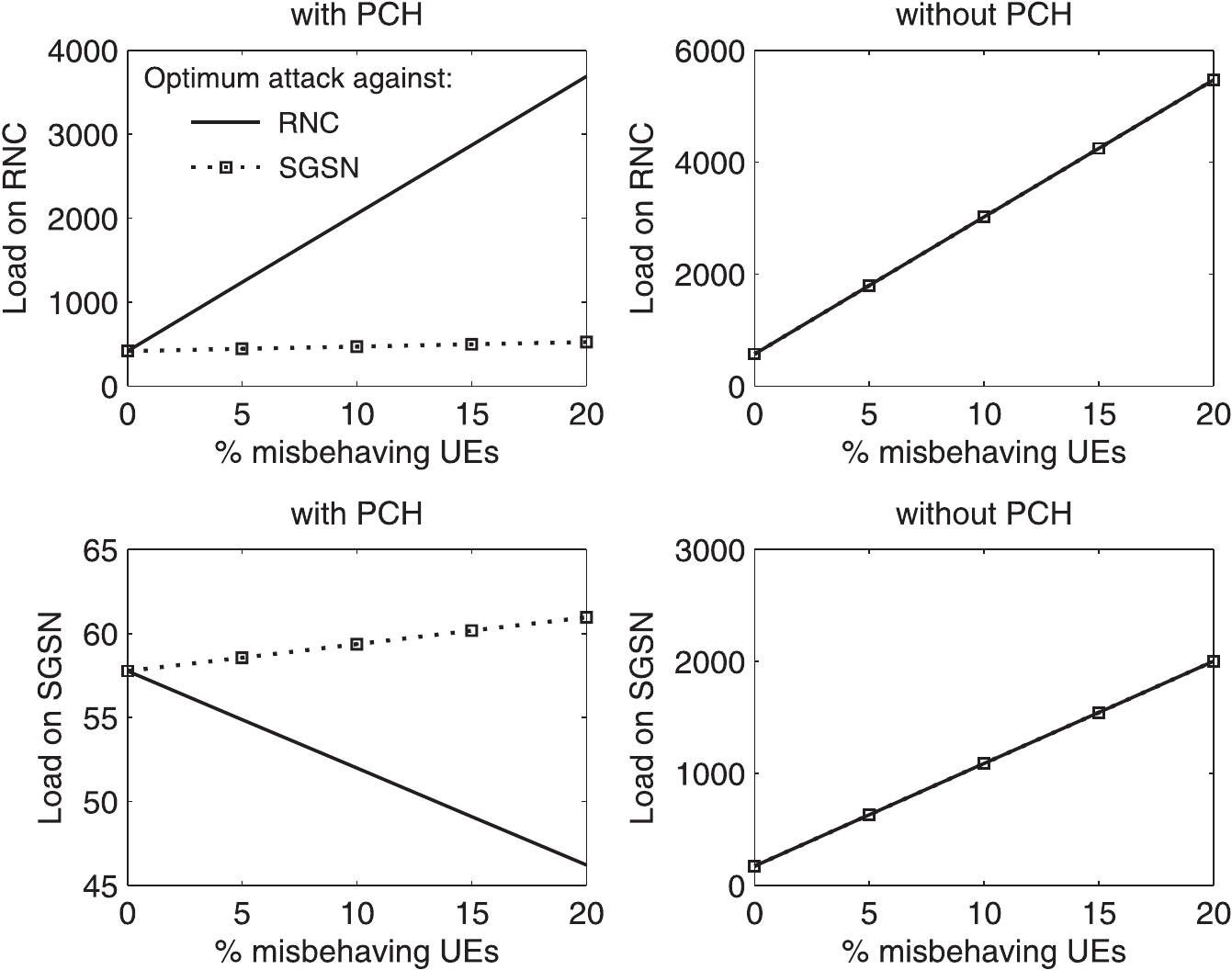}
   \caption{Load on RNC and SGSN versus percentage of mobile devices participating in a storm out of 10,000 users, when $\lambda_L^{-1} = 600, \mu_L^{-1} = 5, \lambda_H^{-1} = 600, \mu_H^{-1} = 180, \tau_H^{-1} = \tau_L^{-1}=5$s and $\tau_P^{-1}=10$~min. When PCH is enabled a storm can cause maximum load on {\em either} the radio or core network, but without PCH both of them could be targeted simultaneously.} \label{fig-storm}
\end{figure}

Finally, we examine in Fig.~\ref{fig-storm} the effect of a signalling storm on the RNC and SGSN when the total number of UEs is 10,000 and the percentage of misbehaving ones is increased from 0 to 20\%.  Comparing the maximum load on the targeted network component and the corresponding load on the other, we see that PCH state prevents a situation where both the RNC and SGSN are simultaneously exposed to worst case loads, which happens when IDLE$\to$ FACH is the bottleneck transition in the radio network (cf. Section~\ref{sec:storm_radio}). In general, the radio network is less sensitive to the choice of the malicious bursts, as long as they are frequent, and thus it is more vulnerable to signalling storms. On the other hand, the load on the core network changes dramatically when the storm is optimised, which may not happen often, making signalling overloads in the SGSN a less likely event. This does not, however, include the effect of complex pricing and business models used by the operator which may exacerbate signalling load in the core network.

\section{Conclusions}

This paper has focused on the  behaviour of a mobile network user with a view to determining network overload in signalling servers and base stations that can result from signalling misbehaviours such as signalling storms. Such misbehaviours can be caused by poorly designed mobile apps, outages in cloud services, large scale malware infections, or malicious network attacks. In the course of this work we have derived a Markov model of user behaviour that can also be exploited in other studies concerning mobile networks as a whole. The Markov model has been solved analytically, and used to derive conditions and parameters for which the signalling misbehaviours can cause the largest damage and which therefore need to be avoided. The analytical results have been illustrated with several numerical examples, and we expect that this work will lead to ideas relating to control algorithms that can adaptively react to network measurements so as to eliminate or mitigate the effect of signalling storms and DoS attacks.

\section*{Acknowledgment}
The authors acknowledge the support of the EU FP7 project NEMESYS, Grant Agreement no. 317888.

\end{document}